\documentclass{elsart}
\usepackage{graphicx}
\usepackage{amsmath}
\usepackage{amssymb}
\usepackage{epsfig}
\usepackage{dcolumn}
\usepackage{bm}
\begin{document}
\begin{frontmatter}

\title{Multifractal spectrum of the phase space related to generalized thermostatistics}
\author{A.I. Olemskoi},
\ead{alex@ufn.ru}
\author{V.O. Kharchenko}
\ead{vasiliy\_ua@ukr.net}
\address{Department of Physical Electronics,
Sumy State University \break 2, Rimskii-Korsakov St., 40007 Sumy, Ukraine}
\date{}

\begin{abstract}
We consider the set of monofractals within a multifractal related to the phase
space being the support of a generalized thermostatistics. The statistical
weight exponent $\tau(q)$ is shown to can be modeled by the hyperbolic tangent
deformed in accordance with both Tsallis and Kaniadakis exponentials whose
using allows one to describe explicitly arbitrary multifractal phase space. The
spectrum function $f(d)$, determining the specific number of monofractals with
reduced dimension $d$, is proved to increases monotonically from minimum value
$f=-1$ at $d=0$ to maximum $f=1$ at $d=1$. The number of monofractals is shown
to increase with growth of the phase space volume at small dimensions $d$ and
falls down in the limit $d\to 1$.
\end{abstract}

\begin{keyword}
Multifractal; exponent of statistical weight; fractal dimension spectrum;
deformed exponential. \PACS 05.45.Df.
\end{keyword}
\end{frontmatter}

\section{Introduction}

Very recently, we have considered a generalized thermostatistics based on a
multifractal phase space, whose dimension $D$ is not equal the number $6N$ of
the degrees of freedom fixed by the number of particles $N$ \cite{Olemskoi}. In
such a case, the behaviour of the complex system is determined by the
dimensionless volume $\gamma=\Gamma/(2\pi\hbar)^{6N}$ of the supporting phase
space where $\hbar$ is Dirac-Planck constant. According to self-similarity
condition, the specific statistical weight of such a system is given by the
power law function \cite{Feder}
\begin{equation}
\varpi_q(\gamma)=\gamma^{qd} \label{1}
\end{equation}
where $q$ is multifractal index, $d\equiv{D}/{6N}\leq 1$ is reduced fractal
dimension. This function should be multiplied by the specific number of
monofractals with dimension $d$
\begin{equation}
\mathcal{N}_d(\gamma)=\gamma^{-f(d)}
\label{2}
\end{equation}
which are contained in multifractal whose spectrum is determined by a function
$f(d)$. As a result, the whole statistical weight, being the multifractal
measure, takes the form
\begin{equation}
w_q(\gamma)\equiv\int\limits^1_0\varpi_q(\gamma)\mathcal{N}_d(\gamma)\rho(d){\rm
d}d=\int\limits^1_0\gamma^{qd-f(d)}\rho(d){\rm d}d \label{3}
\end{equation}
where $\rho(d)$ is a density distribution over dimensions $d$. Using the method
of steepest descent, we arrive at the power law
\begin{equation}
w_q(\gamma)\simeq\gamma^{\tau(q)}\label{4}
\end{equation}
which generalizes the simplest relation (\ref{1}) due to replacement of the
bare fractal dimension $d$ related the index $q=1$ by the multifractal function
\begin{equation}
\tau(q)= qd_q-f(d_q).\label{5}
\end{equation}
Here, the fractal dimension $d_q$ relates to given parameter $q$ to be defined
by the conditions of application of the steepest descent method:
\begin{equation}
\left.\frac{{\rm d}f}{{\rm d}d}\right|_{d=d_q}=q,\quad\left.\frac{{\rm d}^2
f}{{\rm d}d^2}\right|_{d=d_q}<0. \label{6}
\end{equation}

Inserting the statistical weight (\ref{4}) into the deformed entropy
\cite{Naudts,1}
\begin{equation}
S(W)=\int\limits_{\gamma(1)}^{\gamma(W)}\frac{{\rm
d}\gamma}{w(\gamma)}\label{7}
\end{equation}
related to the whole statistical weight $W$ arrives at thermostatistical scheme
governed by the Tsallis formalism of the non-extensive statistics
\cite{Tsallis}. Within such a scheme, the non-extensivity parameter is
determined by the multifractal index (\ref{5}) which monotonically increases,
taking value $\tau=0$ at $q=1$ and $\tau\simeq 1$ at $q\to\infty$. It is
appeared, the multifractal function $\tau(q)$ is reduced to the specific heat
to determine, together with the inverted value ${\bar\tau}(q)\equiv
1/\tau(q)>1$, both statistical distributions and thermodynamic functions of a
complex system \cite{Olemskoi}. In this way, the entropy (\ref{7}) will be
positive definite, continuous, symmetric, expandable, decisive, maximal,
concave, Lesche stable and fulfilling a generalized Pesin-like identity if the
exponent (\ref{5}) varies within the interval $[0,1]$ \cite{Tsallis}.

In this letter, we are aimed to model possible analytical forms of both
multifractal index $\tau(q)$ and related spectrum $f(d)$. In Section 2 we show
that the monotonically increasing function $\tau(q)$ is presented by the
hyperbolic tangent deformed in accordance with both Tsallis and Kaniadakis
procedures whose using allows one to describe explicitly arbitrary multifractal
phase space. Section 3 is devoted to consideration of the multifractal spectrum
$f(d)$ which determines the number of monofractals within the multifractal with
the specific dimension $d$. Section 4 concludes our consideration.

\section{Analytically modeling multifractal}

\begin{figure}[!h]
\centering
\includegraphics[width=80mm]{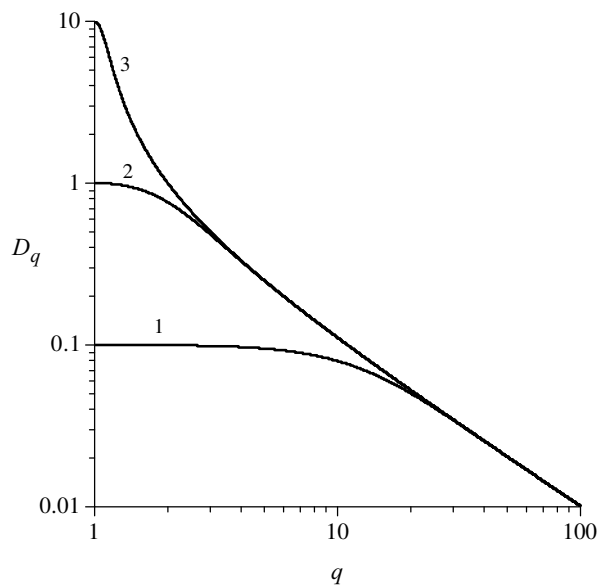}
\caption{Multifractal dimension function related to the dependence (\ref{8})
(curves 1, 2, 3 correspond to $\mathcal D=0.1, 1.0, 10$).}\label{Dq(q)_D}
\end{figure}

As the simplest case, we take the function $\tau(q)$ in the form
\begin{equation}
\tau=\tanh\left[\mathcal D(q-1)\right],\label{8}
\end{equation}
being determined by parameter $\mathcal D>0$ and argument $q\in[1,\infty)$.
According to Fig.\ref{Dq(q)_D}, related multifractal dimension function
\cite{Feder}
\begin{equation}
D_q=\frac{\tau(q)}{q-1} \label{9}
\end{equation}
monotonically decreases from maximum value $D_0=\mathcal D$ to $D_\infty=0$
with $q$ increase. However, maximum value of the fractal dimension $D_q$ is
fixed by the magnitude $D_0=1$, so that one should put $\mathcal D=1$ in the
dependence (\ref{8}). As a result, it takes quite trivial form.

As the $\tau(q)$ function increases monotonically within narrow interval
$[0,1]$, one has a scanty choice of its modeling. One of such possibilities is
opened with using deformation of the hyperbolic tangent (\ref{8}) at $\mathcal
D=1$. At present, two analytical procedures are extensively popularized. The
first of these is based on the Tsallis exponential \cite{Ts}
\begin{eqnarray}
\exp_{\kappa}(x)\equiv\left\{
\begin{array}{ll}
\left(1+\kappa x\right)^{1/\kappa}\quad\ {\rm at}\quad 1+\kappa x>0,\\ 0\ \
\qquad\quad\qquad\qquad\quad{\rm otherwise}
\end{array} \right.
\label{10}
\end{eqnarray}
where deformation parameter $\kappa$ takes positive values. The second
procedure has been proposed by Kaniadakis \cite{Kaniadakis} to determine the
deformed exponential
\begin{equation}
\exp_{\kappa}(x)\equiv\left(\kappa x+\sqrt{1+\kappa^2 x^2}\right)^{1/\kappa}.
\label{11}
\end{equation}
With using these definitions, the deformed tangent (\ref{8}) takes the form
\begin{equation}
\tau_{\kappa}(q)=\tanh_{\kappa}(q-1)\equiv\frac{\exp_{\kappa}(q-1)-
\exp_{\kappa}(1-q)}{\exp_{\kappa}(q-1)+\exp_{\kappa}(1-q)}\label{12}
\end{equation}
where the multifractal index $q$ varies within the domain $[1,\infty)$.

\begin{figure}[!h]
\centering
\includegraphics[width=80mm]{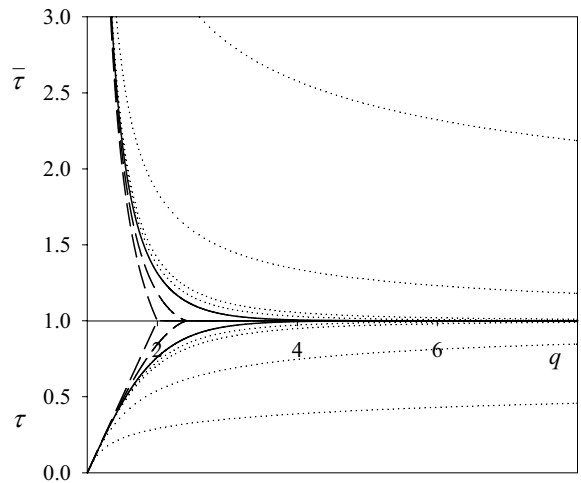}
\caption{The $q$-dependencies of the multifractal index $\tau$ and its inverted
value $\bar\tau=1/\tau$ (solid line corresponds to $\kappa=0$, dashed curves
relate to the Tsallis deformation with $\kappa=0.7, 1$; dotted lines correspond
to the Kaniadakis one at $\kappa=0.7, 1, 3, 10$).}\label{tau(q)}
\end{figure}

The $q$-dependencies of the multifractal index $\tau(q)$ and its inverted value
$\bar\tau(q)=1/\tau(q)$ are shown in Fig.\ref{tau(q)} at different magnitudes
of both Tsallis and Kaniadakis deformation parameters $\kappa$. It is
principally important, the first of these deformations arrives at more fast
variations of both indices $\tau(q)$ and $\bar\tau(q)$ in comparison with
non-deformed hyperbolic tangent $\tau_0=\tanh_0(q-1)$, whereas the Kaniadakis
deformation slows down these variations with $\kappa$ increase.

Characteristic peculiarity of the Tsallis deformation consists in breaking
dependencies $\tau(q)$, $\bar\tau(q)$ in the point $q_0=(1+\kappa)/\kappa$
where the second terms in both numerator and denominator of the definition
(\ref{12}) take the zero value. As a result, the multifractal index (\ref{12})
obtains the asymptotics
\begin{eqnarray}
\tau_{\kappa}^{(Ts)}\simeq\left\{
\begin{array}{ll}
\left(q-1\right)-\frac{1-\kappa^2}{3}\left(q-1\right)^3 \quad\qquad {\rm
at}\quad\ 0<q-1\ll 1,\\
1-2\left(\kappa/2\right)^{1/\kappa}\left(q_0-q\right)^{1/\kappa}\quad\quad {\rm
at}\quad 0<q_0-q\ll q_0.
\end{array} \right.
\label{13}
\end{eqnarray}
For $\kappa=1$ the dependence $\tau_1^{(Ts)}(q)$ takes the simplest form:
$\tau_1^{(Ts)}=q-1$ at $1\leq q\leq 2$, and $\tau_1^{(Ts)}=1$ at $q>2$. It is
worthwhile to notice, the Tsallis deformation parameter can not take values
$\kappa>1$ because these relate to fractal dimensions $D_q>1$ at $q\ne 0$.

At the Kaniadakis deformation, the multifractal index $\tau_{\kappa}(q)$ varies
smoothly to be characterized by the following asymptotics:
\begin{eqnarray}
\tau_{\kappa}^{(K)}\simeq\left\{
\begin{array}{ll}
\left(q-1\right)-\frac{2+\kappa^2}{6}\left(q-1\right)^3 \qquad {\rm at}\quad \
0<q-1\ll 1,\\ 1-2\left[2\kappa(q-1)\right]^{-2/\kappa}\quad\quad\quad {\rm
at}\quad\quad \kappa(q-1)\gg 1.
\end{array} \right.
\label{14}
\end{eqnarray}
In contrast to the Tsallis case, here the deformation parameter can take
arbitrary values to give the simplest dependence
$\tau_1^{(K)}=\frac{q-1}{\sqrt{1+(q-1)^2}}$ at $\kappa=1$.

\begin{figure}[!h]
\centering
\includegraphics[width=60mm]{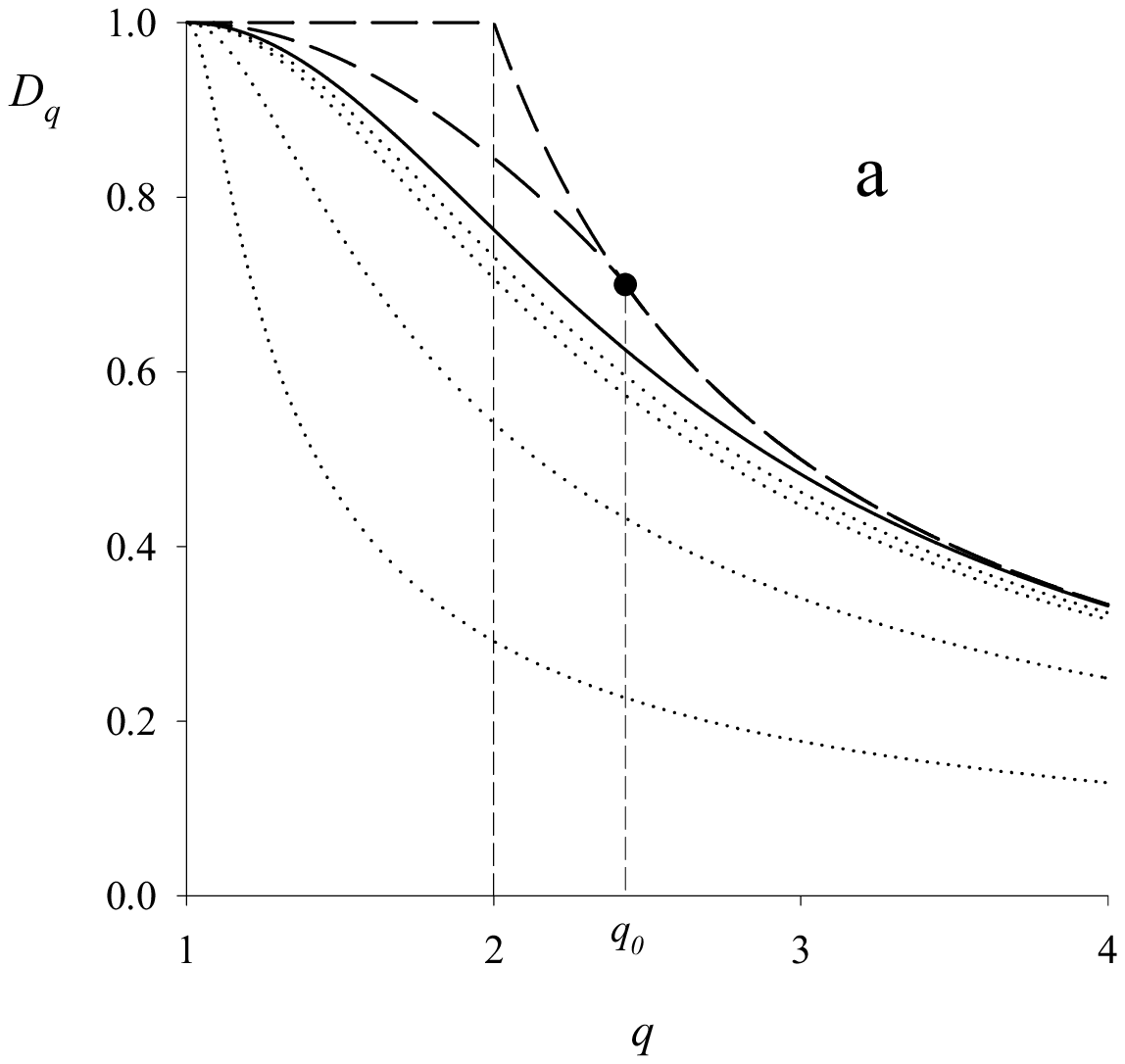}
\hspace{0.5cm}\includegraphics[width=60mm]{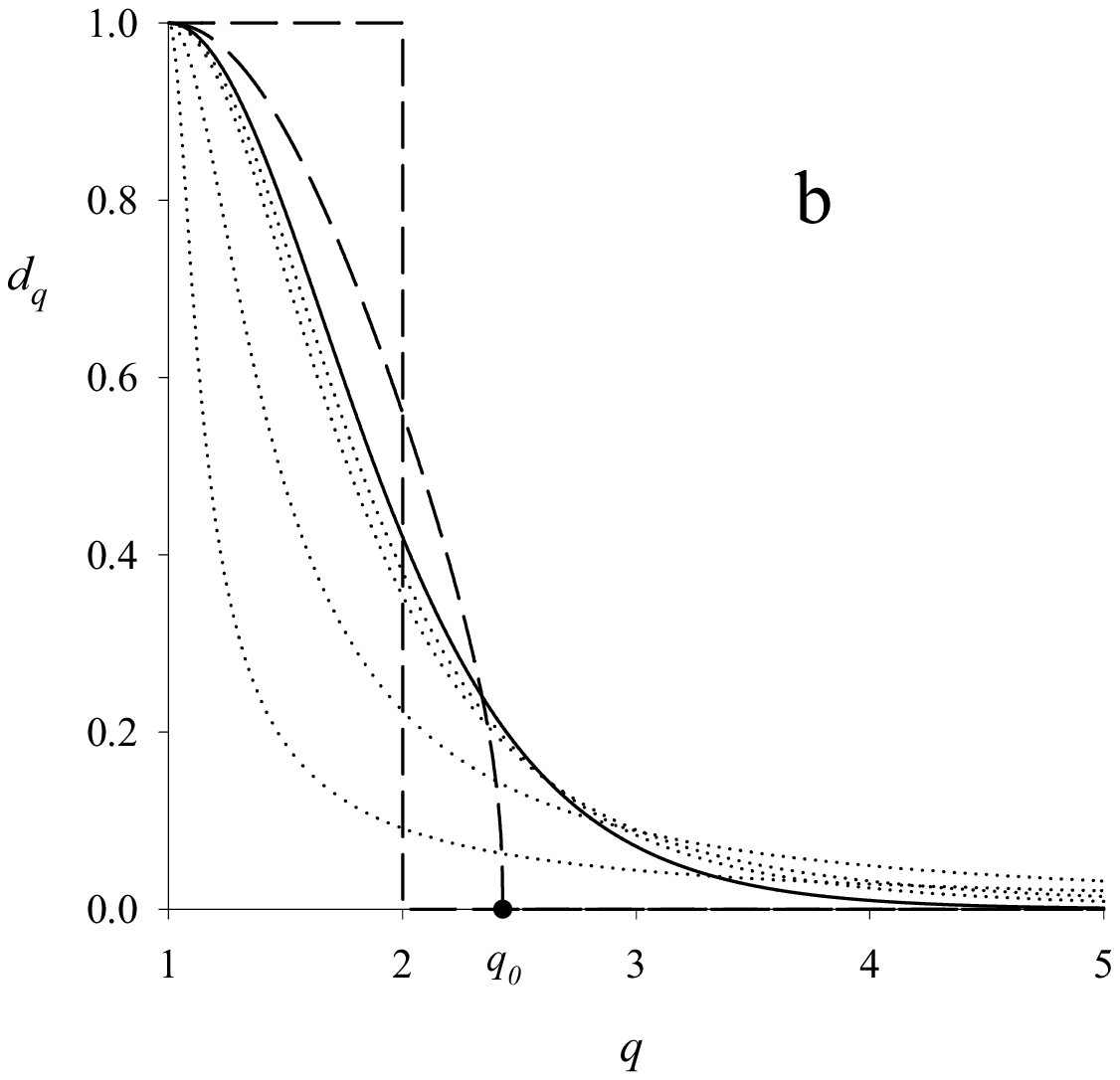}
 \caption{Spectra of the
fractal (a) and specific (b) dimensions of the phase space (the choice of the
deformation parameters is the same as in Fig.\ref{tau(q)}).}\label{d_q}
\end{figure}

As a function of the index $q$, the fractal dimension (\ref{9}) falls down
monotonically as shown in Fig.\ref{d_q}a. According to Eqs.(\ref{13}), at the
Tsallis deformation, one has broken dependence $D(q)$, being characterized by
the asymptotics
\begin{eqnarray}
D_q^{(Ts)}\simeq\left\{
\begin{array}{ll}
1-\frac{1-\kappa^2}{3}\left(q-1\right)^2 \quad\quad\quad\qquad {\rm at}\quad\ \
0<q-1\ll 1,\\
\frac{1}{q-1}-2\left(\kappa/2\right)^{1/\kappa}\frac{\left(q_0-q\right)^{1/\kappa}}
{q-1}\quad\quad {\rm at}\quad 0<q_0-q\ll q_0.
\end{array} \right.
\label{15}
\end{eqnarray}
In the limit case $\kappa=1$, the phase space is smooth $(D_q^{(Ts)}=1)$ within
the interval $1\leq q \leq 2$. For the Kaniadakis deformation, the fractal
dimension $D_q^{(K)}$ is given by smoothly falling down curve whose slope
increase with the deformation parameter growth. According to Eqs.(\ref{14}), in
this case, one has the asymptotics
\begin{eqnarray}
D_q^{(K)}\simeq\left\{
\begin{array}{ll}
1-\frac{2+\kappa^2}{6}\left(q-1\right)^2 \quad\quad\qquad\qquad\qquad {\rm
at}\quad \ 0<q-1\ll 1,\\
(q-1)^{-1}-2\left(2\kappa\right)^{-2/\kappa}(q-1)^{-(\kappa+2)/\kappa}\ \ {\rm
at}\ \ \kappa(q-1)\gg 1.
\end{array} \right.
\label{16}
\end{eqnarray}
At $\kappa=1$, the typical dependence reads
$D_q^{(K)}=\left[1+(q-1)^2\right]^{-1/2}$.

\section{Multifractal spectrum}

At given multifractal index $\tau(q)$, the spectrum function $f(d)$ is defined
by inverted Legendre transformation (\ref{5}) where the specific dimension
reads
\begin{equation}
d_q=\frac{{\rm d}\tau}{{\rm d}q}. \label{17}
\end{equation}
As shows Fig.\ref{d_q}b, the dependence $d(q)$ has monotonically falling down
form to take the value $d_q=0$ at $q>q_0\equiv(1+\kappa)/\kappa$ for the
Tsallis deformation. In this case, asymptotical behaviour is characterized by
Eqs.(\ref{13}), according to which one obtains
\begin{eqnarray}
d_q^{(Ts)}\simeq\left\{
\begin{array}{ll}
1-(1-\kappa^2)\left(q-1\right)^2 \quad\quad\quad\quad\ \ {\rm at}\quad\
0<q-1\ll 1,\\
\left(\kappa/2\right)^{(1-\kappa)/\kappa}\left(q_0-q\right)^{(1-\kappa)/\kappa}\qquad
{\rm at}\quad 0<q_0-q\ll q_0.
\end{array} \right.
\label{18}
\end{eqnarray}
In the limit $\kappa\to 1$, the dependence $d^{(Ts)}(q)$ takes the step-like
form being $d_q=1$ within the interval $1\leq q\leq 2$ and $d_q=0$ otherwise.

For the Kaniadakis deformation, Eqs.(\ref{14}) arrive at the asymptotics
\begin{eqnarray}
d_q^{(K)}\simeq\left\{
\begin{array}{ll}
1-\frac{2+\kappa^2}{2}\left(q-1\right)^2 \quad\quad\qquad\qquad {\rm at}\quad \
0<q-1\ll 1,\\
2^{2(\kappa-1)/\kappa}\left[\kappa(q-1)\right]^{-(2+\kappa)/\kappa}\quad\quad
{\rm at}\quad\quad \kappa(q-1)\gg 1.
\end{array} \right.
\label{19}
\end{eqnarray}
The typical behaviour is presented by the dependence
$d_q^{(K)}=\left[1+(q-1)^2\right]^{-3/2}$ related to $\kappa=1$.

The multifractal spectrum is defined by the equality
\begin{equation}
f(d)=dq_d-\tau(q_d)\label{5a}
\end{equation}
being conjugated Eq.(\ref{5}). Here, the specific multifractal index $q_d$ is
determined by Eq.(\ref{17}) which arrives at the limit relations (\ref{18}),
(\ref{19}). With its using, one obtains the asymptotics
\begin{eqnarray}
f^{(Ts)}\simeq\left\{
\begin{array}{ll}
d-\frac{2}{3}(1-\kappa^2)^{-1/2}(1-d)^{3/2} \quad\quad\quad\ \ {\rm at}\quad\
0<1-d\ll 1,\\ -\left[1+2(1/\kappa-1)d^{1/(1-\kappa)}\right]+(1+1/\kappa)d\qquad
{\rm at}\quad\ d\ll 1
\end{array}
\right. \label{20}
\end{eqnarray}
for the Tsallis deformation, and the relations
\begin{eqnarray}
f^{(K)}\simeq\left\{
\begin{array}{ll}
d-\frac{2}{3}\left(1+\frac{\kappa^2}{2}\right)^{-1/2}(1-d)^{3/2} \quad\quad\ \
{\rm at}\quad 0<1-d\ll 1,\\
-\left[1-2^{(\kappa-4)/(2+\kappa)}(1+2/\kappa)d^{2/(2+\kappa)}\right]+d\quad\
{\rm at}\quad d\ll 1,
\end{array} \right.
\label{21}
\end{eqnarray}
characterizing the Kaniadakis deformation.

\begin{figure}[!h]
\centering
\includegraphics[width=80mm]{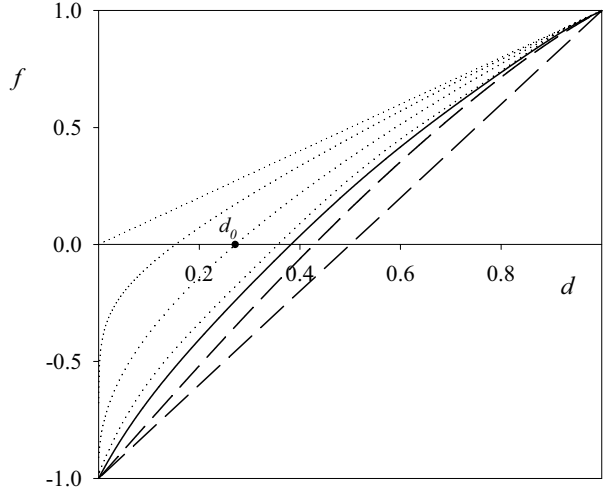}
\caption{Spectrum function of the multifractal phase space (solid line
corresponds to $\kappa=0$, dashed curves relate to the Tsallis deformation with
$\kappa=0.7, 1$; dotted lines correspond to the Kaniadakis one at $\kappa=1, 3,
10,\infty$).}\label{f(d)}
\end{figure}

As shows Fig.\ref{f(d)}, for finite deformation parameters $\kappa<\infty$,
a spectrum function increases monotonically, taking the minimum value $f=-1$ at
$d=0$ and maximum $f=1$ at $d=1$. By this, the derivative $f'\equiv{\rm
d}f/{\rm d}d$ equals $f'(0)=\infty$ on the left boundary and $f'(1)=1$ on the
right one. It is characteristically, the whole set of the spectrum functions is
bounded by the limit dependencies $f^{(Ts)}=2d-1$ and $f^{(K)}=d$, the first of
which relates to limit magnitude of the Tsallis deformation parameter
$\kappa=1$ whereas the second corresponds to the Kaniadakis limit
$\kappa=\infty$. Typical form of the spectrum function is presented by the
dependencies
\begin{eqnarray}
f^{(K)}=\left\{
\begin{array}{ll}
-d\ln\left(\frac{\sqrt{d}}{1+\sqrt{1-d}}\right)+\left(d-\sqrt{1-d}\right)\quad
{\rm at}\quad
\kappa=0,\\-\left(1-d^{2/3}\right)^{3/2}+d\quad\quad\quad\quad\quad\quad\ \
{\rm at}\quad\kappa=1.
\end{array} \right.
\label{22}
\end{eqnarray}

\begin{figure}[!h]
\centering
\includegraphics[width=70mm]{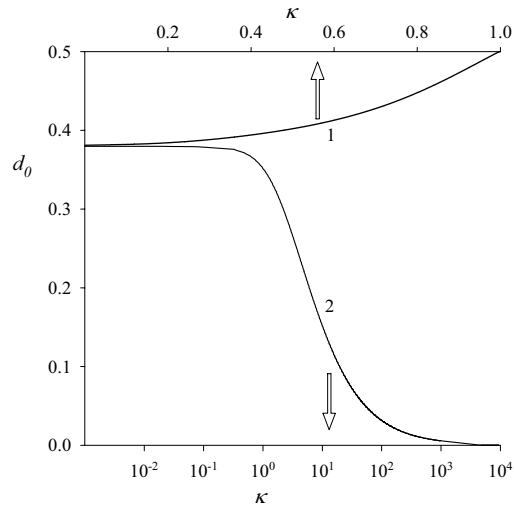}
\caption{Dimension $d_0$ related to the condition $f(d)=0$ as function of the
parameter $\kappa$ (curve 1 corresponds to the Tsallis deformation, curve 2 --
to the Kaniadakis one; positive values $f(d)$ relate to the domain
$d>d_0$).}\label{d0(k)}
\end{figure}

It may seem, at first glance, that negative values of the spectrum function
$f(d)$ has not a physical meaning. To clear up this question, let us take the
set of monofractals with the reduced dimension $d=0$. Obviously, such
monofractals relate to the whole set of the phase space points, whose number
equals the dimensionless volume $\gamma$. Just such result gives the definition
(\ref{2}) in the point $d=0$ where $f=-1$. On the other hand, in opposite case
$d=1$ where $f=1$, we obtain the very thing the number of monofractals with
volume $\gamma$ equals $\mathcal{N}_1=\gamma^{-1}$ to give one multifractal of
the same volume $\gamma$. In this way, a single monofractal is contained in the
multifractal at condition $f(d)=0$ which takes place at the reduced dimension
$d_0$ whose dependence on the deformation parameter $\kappa$ is shown in
Fig.\ref{d0(k)}.

\begin{figure}[!h]
\centering \raisebox{30mm}{\tiny{$\mathcal{N}$}}
\includegraphics[width=60mm]{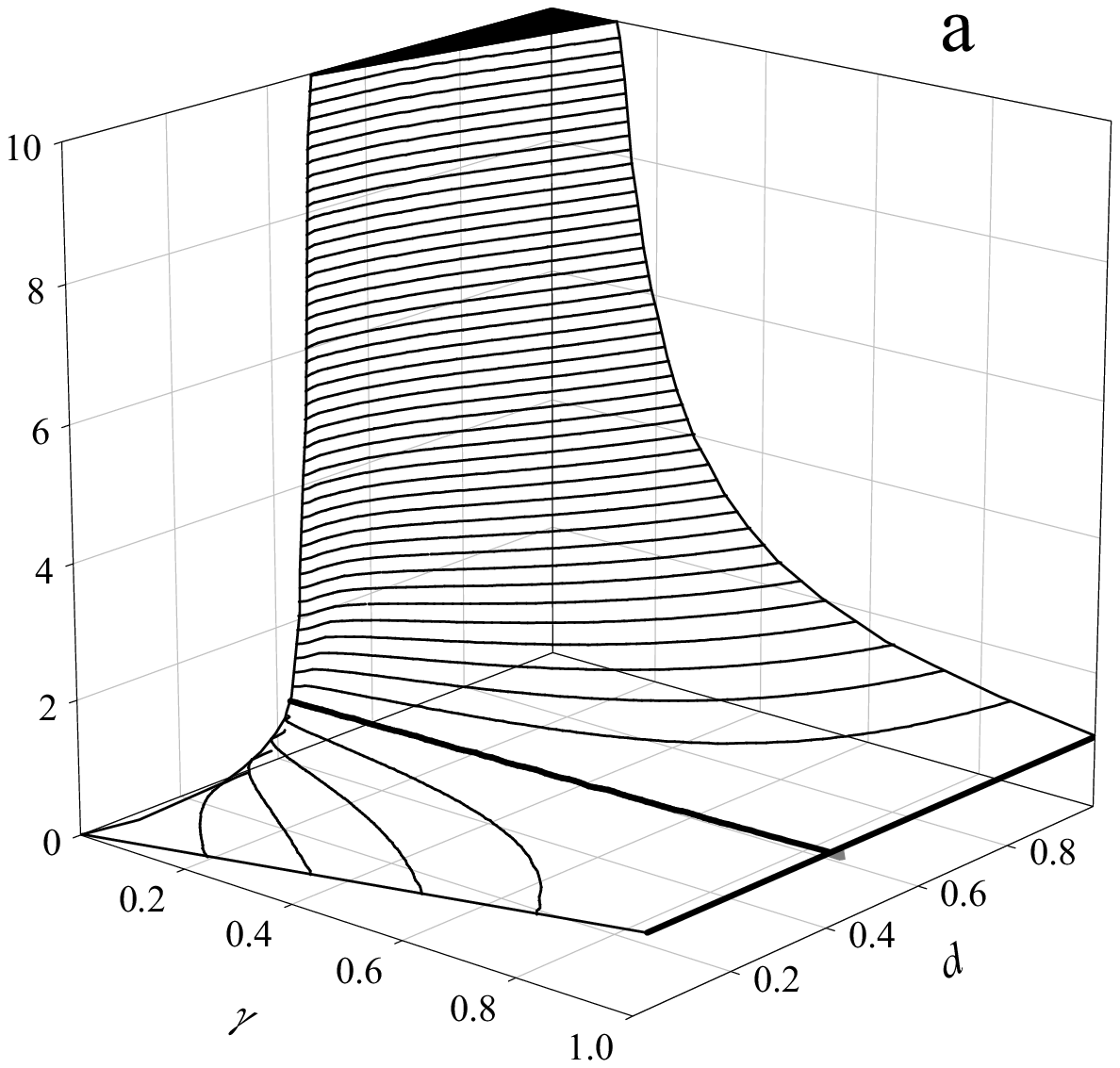}\hspace{1.0cm}
\centering \raisebox{30mm}{\tiny{$\mathcal{N}$}}
\includegraphics[width=60mm]{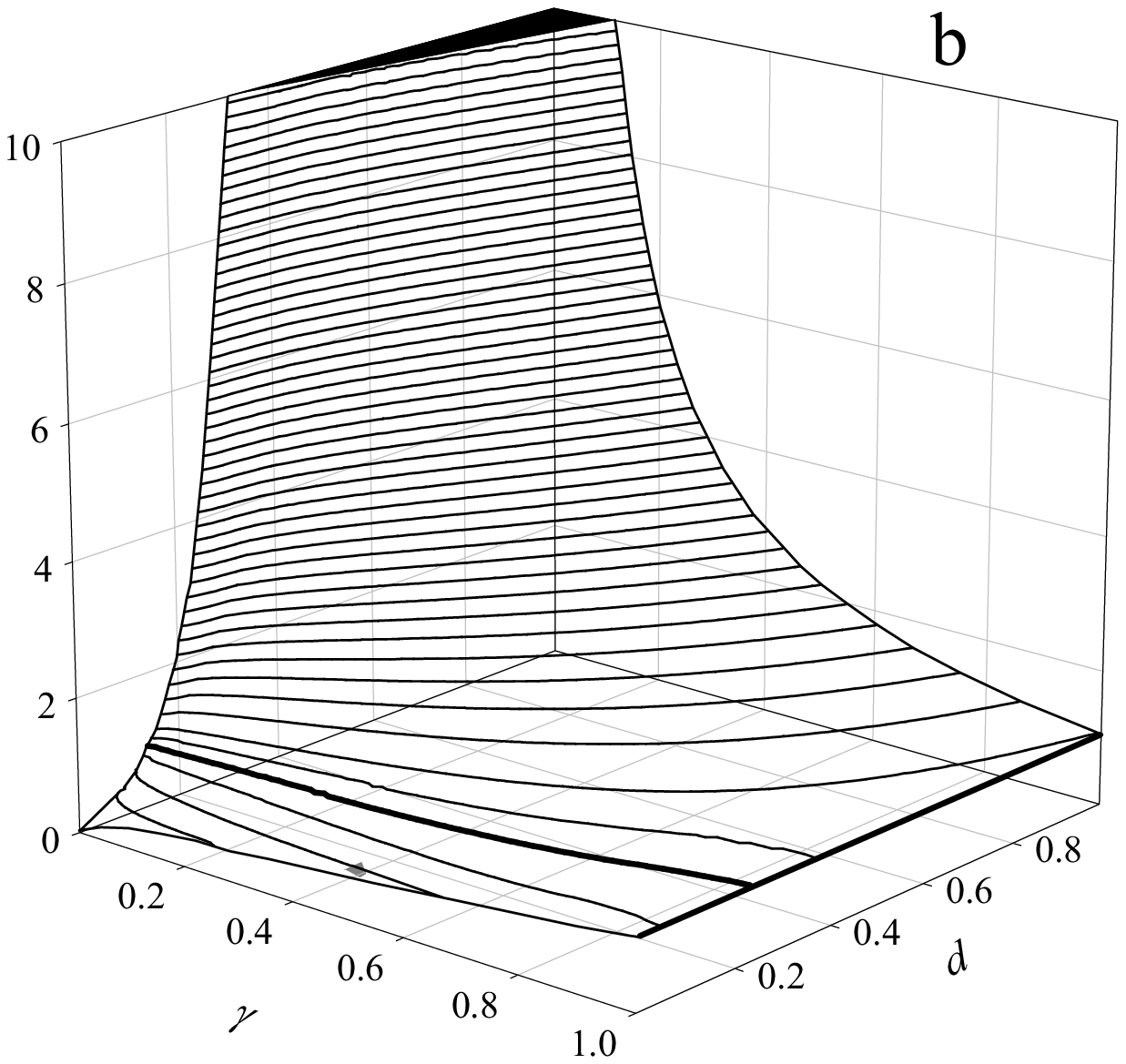}
\caption{The number $\mathcal{N}$ of monofractals within the phase space volume
$\gamma$ related to the specific dimension $d$ at the deformation parameters
$\kappa=0$ (a) and $\kappa=10$ (b) (different levels $\mathcal{N}={\rm const}$
are shown with thin curves, thick lines relate to
$\mathcal{N}=1$).}\label{N(k)}
\end{figure}

The dependence of the number $\mathcal{N}$ of monofractals containing in the
phase space volume $\gamma$ related to the multifractal with the specific
dimension $d$ is shown in Fig.\ref{N(k)}. It is seen, the number
$\mathcal{N}$ increases with the $\gamma$ volume growth at small dimensions
$d$, whereas in the limit $d\to 1$ the dependence $\mathcal{N}(\gamma)$ becomes
falling down to give infinitely increasing numbers $\mathcal{N}$ at $\gamma\to
0$. The speed of such increase growths monotonically with both decrease of the
Tsallis deformation parameter $\kappa$ and increase of the Kaniadakis one.

\section{Conclusions}

The whole set of monofractals within a multifractal related to the phase space,
which gives the support of a generalized thermostatistics, is modeled by the
exponent $\tau(q)$ that determines the statistical weight (\ref{4}) at given
volume $\gamma$. To be the entropy (\ref{7}) concave, Lesche stable et cetera,
the exponent $\tau(q)$ should be a function, monotonically increasing within
the interval $[0,1]$ at multifractal exponent variation within the domain
$[1,\infty)$. The simplest case of such a function gives the hyperbolic tangent
$\tau=\tanh(q-1)$ whose deformation (\ref{12}) defined in accordance with both
Tsallis and Kaniadakis exponentials (\ref{10}), (\ref{11}) allows one to
describe explicitly arbitrary multifractal phase space. In this way, the
Tsallis deformation arrives at more fast variations of the statistical weight
index $\tau(q)$ in comparison with non-deformed hyperbolic tangent, whereas the
Kaniadakis one slows down these variations with increasing the deformation
parameter $\kappa$. All possible dependencies $\tau(q)$ are bounded from above
by the linear function $\tau^{(Ts)}=q-1$ at $q\in [1,2]$ which is transformed
into the constant $\tau=1$ at $q>2$. This dependence relates to the smooth
phase space within the Tsallis interval $q\in [1,2]$.

The dependence (\ref{2}) of the number of monofractals within the phase space
volume $\gamma$ related to the multifractal with the specific dimension $d$ is
determined by the spectrum function $f(d)$. This function increases
monotonically, taking the minimum value $f=-1$ at $d=0$ and maximum $f=1$ at
$d=1$; by this, its derivative equals $f'(0)=\infty$ on the left boundary and
$f'(1)=1$ on the right one. The whole set of the spectrum functions is bounded
by the limit dependencies $f^{(Ts)}=2d-1$ and $f^{(K)}=d$, the first of which
relates to limit magnitude of the Tsallis deformation parameter $\kappa=1$ and
the second one corresponds to the Kaniadakis limit $\kappa=\infty$. The number
of monofractals within the multifractal increases with the $\gamma$ volume
growth at small dimensions $d$ and falls down in the limits $d\to 1$ to give
infinitely increasing at $\gamma\to 0$.

\end{document}